%% file: charm2015_Purohit.tex
\pdfoutput=1
\documentclass[12pt]{article}
\include{packages}

\def\Dbar    {\kern 0.2em\overline{\kern -0.2em D}{}\xspace}
\def\Bbar    {\kern 0.18em\overline{\kern -0.18em B}{}\xspace}
\def\babar{\mbox{\slshape B\kern-0.1em{\small A}\kern-0.1em
    B\kern-0.1em{\small A\kern-0.2em R}}\xspace}
\def\modei{$B^0 \to D^- D^0 K^+$}
\def\modexi{$B^+\to \overline{D}^0 D^0 K^+$}

\def\Dz{D^0}
\def\Km    {\ensuremath{K^-}\xspace}
\def\pip   {\ensuremath{\pi^+}\xspace}
\def\piz   {\ensuremath{\pi^0}\xspace}
\def\Dp      {\ensuremath{D^+}\xspace}
\def\pim   {\ensuremath{\pi^-}\xspace}
\def\de    {\mbox{$\Delta E$}\xspace}
\def\mes        {\mbox{$m_{\rm ES}$}\xspace}
\def\PsiRes {\ensuremath{\psi(3770)}\xspace}
\def\ndof {\ensuremath{n_{\mathrm{dof}}}}

\newcommand{\jprBase}        {Phys.\ Rev.\xspace}
\newcommand{\jprd}      [1]  {\jprBase\ D~{\bf #1}}
\newcommand\CellTop{\rule{0pt}{2.35ex}}
\newcommand\CellTopTwo{\rule{0pt}{2.8ex}}
\newcommand{\mevcc}{\ensuremath{{\mathrm{\,Me\kern -0.1em V\!/}c^2}}\xspace}
\newcommand{\DsTwo}{\ensuremath{D^*_{s2}(2573)^+}\xspace}
\newcommand{\gevcc}{\ensuremath{{\mathrm{\,Ge\kern -0.1em V\!/}c^2}}\xspace}
\newcommand{\BToDDK}{\ensuremath{B\to \Dbar^{(*)} D^{(*)} K}\xspace}
\newcommand{\DsJ}[1]{\ensuremath{D_{sJ}^*(#1)^+}\xspace}
\newcommand{\DsOne}{\ensuremath{D^{*}_{s1}(2700)^+}\xspace}

\newcommand\pubdate{May 20, 2015}

\def\usc{Department of Physics and Astronomy\\
University of South Carolina, Columbia, SC 29201, USA}
\def\support{\footnote{Work supported by the U.S. Department of Energy.}}

\def\Title#1{\begin{center} {\Large #1 } \end{center}}
\def\Author#1{\begin{center}{ \sc #1} \end{center}}
\def\Address#1{\begin{center}{ \it #1} \end{center}}

\newcommand\pubblock{\rightline{\begin{tabular}{l} \\
         \pubdate  \end{tabular}}}
\newenvironment{Abstract}{\begin{quotation}  }{\end{quotation}}
\newenvironment{Presented}{\begin{quotation} \begin{center} 
             PRESENTED AT\end{center}\bigskip 
      \begin{center}\begin{large}}{\end{large}\end{center} \end{quotation}}
\def\Acknowledgements{\bigskip  \bigskip \begin{center} \begin{large}
             \bf ACKNOWLEDGEMENTS \end{large}\end{center}}

\input econfmacros.tex

\begin{document}
\begin{titlepage}
\pubblock

\vfill
\Title{Dalitz plot analysis of $B\to DDK$ decays}
\vfill
\Author{Milind V. Purohit\support}
\Address{\usc}
\vfill
\begin{Abstract}
    We present Dalitz plot analyses for the decays of $B$ mesons to
  $D^-D^0K^+$ and $\Dbar^0D^0K^+$. [Charge conjugate reactions are
    implicitly assumed throughout.] We report the observation of
  the \DsOne resonance in these two channels and obtain
  measurements of the mass $M(D^*_{s1}(2700)^+) = 2699^{+14}_{-7}$
  MeV/$c^2$ and of the width $\Gamma(D^*_{s1}(2700)^+) =
  127^{+24}_{-19}$ MeV, including statistical and systematic
  uncertainties. In addition, we observe an enhancement in the $D^0K^+$
  invariant mass around 2350--2500 MeV/$c^2$ in both decays $B^0 \to
  D^-D^0K^+$ and $B^+ \to \Dbar^0D^0K^+$, which we are not able to
  interpret. The results are based on 429 $fb^{-1}$ of data containing
  $471\times 10^6$ $B\Bbar$ pairs collected at the $\Upsilon(4S)$
  resonance with the \babar detector at the SLAC National Accelerator
  Laboratory.
\end{Abstract}
\vfill
\begin{Presented}
The 7th International Workshop on Charm Physics (CHARM 2015)\\
Detroit, MI, 18-22 May, 2015
\end{Presented}
\vfill
\end{titlepage}
\def\thefootnote{\fnsymbol{footnote}}
\setcounter{footnote}{0}
%

\section{Introduction}
\label{sec:intro}

\noindent
We analyze $B$ decays to $DDK$ final states via Dalitz plots in order to
measure the mass and width of the \DsJ{2700}. This is the first time that
Dalitz plot analyses have been performed for $B\to DDK$ decays.

The data analyzed here were recorded by the \babar\ detector at the
PEP-II asymmetric-energy $e^+e^-$ storage ring operating at the SLAC
National Accelerator Laboratory. This analysis uses the complete
\babar\ data sample collected at the $\Upsilon(4S)$ resonance
corresponding to an integrated luminosity of
429~fb$^{-1}$~\cite{ref:lumi}. The \babar\ detector is described in
detail elsewhere~\cite{ref:nim}.  Our Monte Carlo simulation uses
\textsc{EVTGEN}~\cite{ref:evtgen} to model the kinematics of $B$ mesons
and \textsc{JETSET}~\cite{ref:jetset} to model continuum processes,
$e^+e^-\to q\overline{q}$ ($q=u,d,s,c$). The \babar\ detector and its
response to particle interactions are modeled using the
\textsc{GEANT4}~\cite{ref:geant4} simulation package.

\section{Data set and selection}
\label{sec:selection}

\noindent
The selection and reconstruction of \modei and \modexi, along with 20
other \BToDDK modes, is described in Ref.~\cite{ref:DDKBF}. We
reconstruct $D$ mesons in the modes $\Dz\to \Km\pip$, $\Km\pip\piz$,
$\Km\pip\pim\pip$, and $\Dp\to \Km\pip\pip$. For the mode \modexi, at
least one of the $\Dz$ mesons is required to decay to $\Km \pip$. We use
the beam-energy-substituted mass ($m_{ES}$) and $\de$, the difference
between the reconstructed energy of the $B$ candidate and the beam
energy in the $e^+e^-$ center-of-mass frame to assist in identifying
signal and in case of multiple candidates per event we use the latter to
make a selection.  Finally, we keep only events with $|\de|<10-14$ MeV
depending on the $D$ final state~\cite{ref:DDKBF}.

We fit the \mes distributions, as described in detail in
Ref.~\cite{ref:DDKBF} to obtain signal yields; however, since there are
22 \BToDDK modes that can cross-feed each other an iterative procedure
is employed to obtain $635 \pm 47$ and $901 \pm 54$ signal events for
\modei and \modexi, respectively~\cite{ref:DDKBF}. For the Dalitz
analyses we use the $5.275 < \mes < 5.284 \gevcc$ region to obtain a
total of 1470 events with a signal purity of $(38.6 \pm 2.8 \pm 2.1)\%$
for \modei and 1894 events with a signal purity of $(41.6 \pm 2.5 \pm
3.1)\%$ for \modexi, where the first uncertainties are statistical and
the second systematic.

\begin{figure}[htb]
 \centering
  \epsfig{file=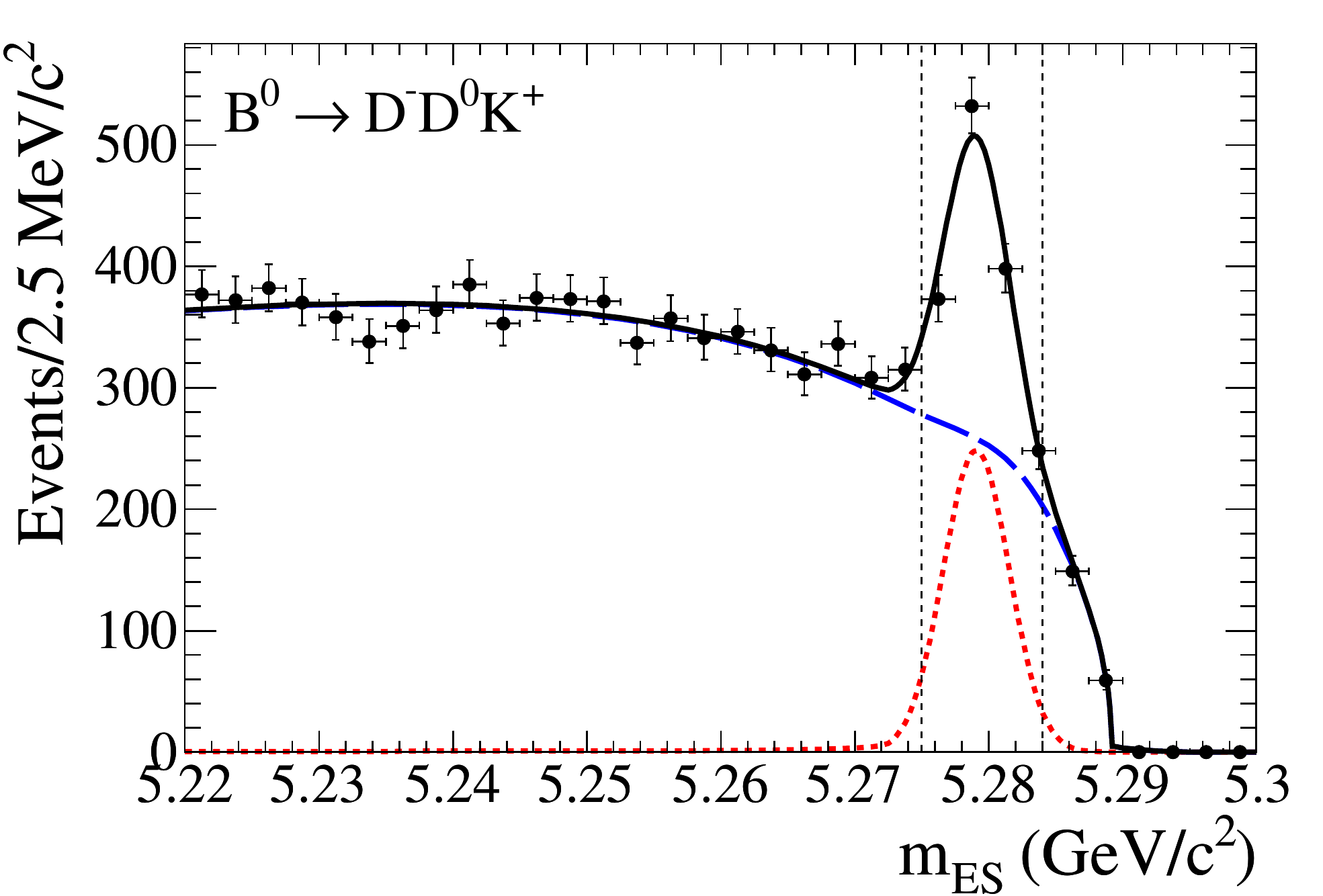,width=0.45\textwidth}
  \epsfig{file=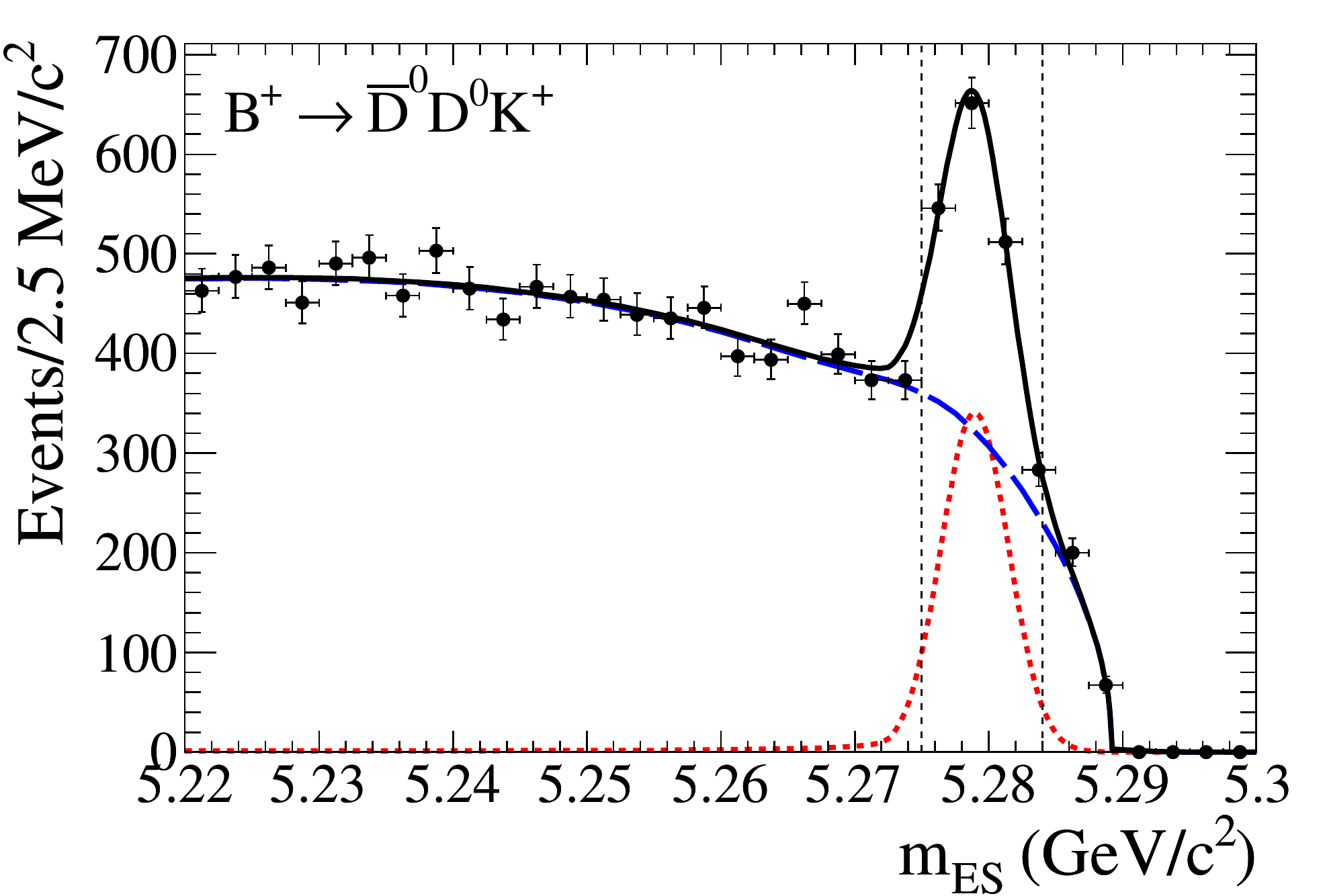,width=0.45\textwidth}
 \caption{Fits of the \mes\ data distributions~\cite{ref:DDKBF} for the
   modes \modei (left) and \modexi (right).}
 \label{fig:mes}
\end{figure}

\begin{figure}[htb]
 \centering
  \epsfig{file=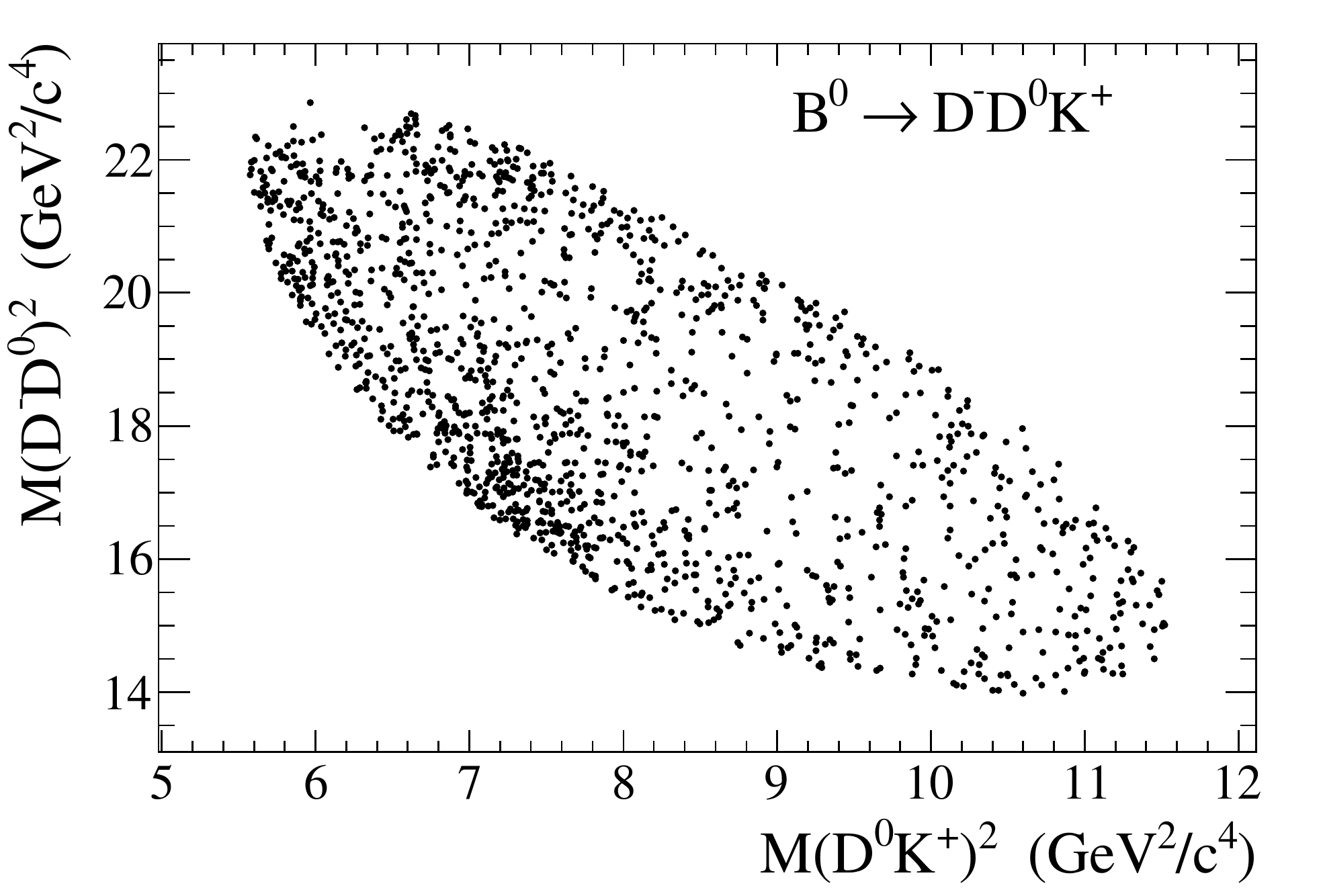,width=0.45\textwidth}
  \epsfig{file=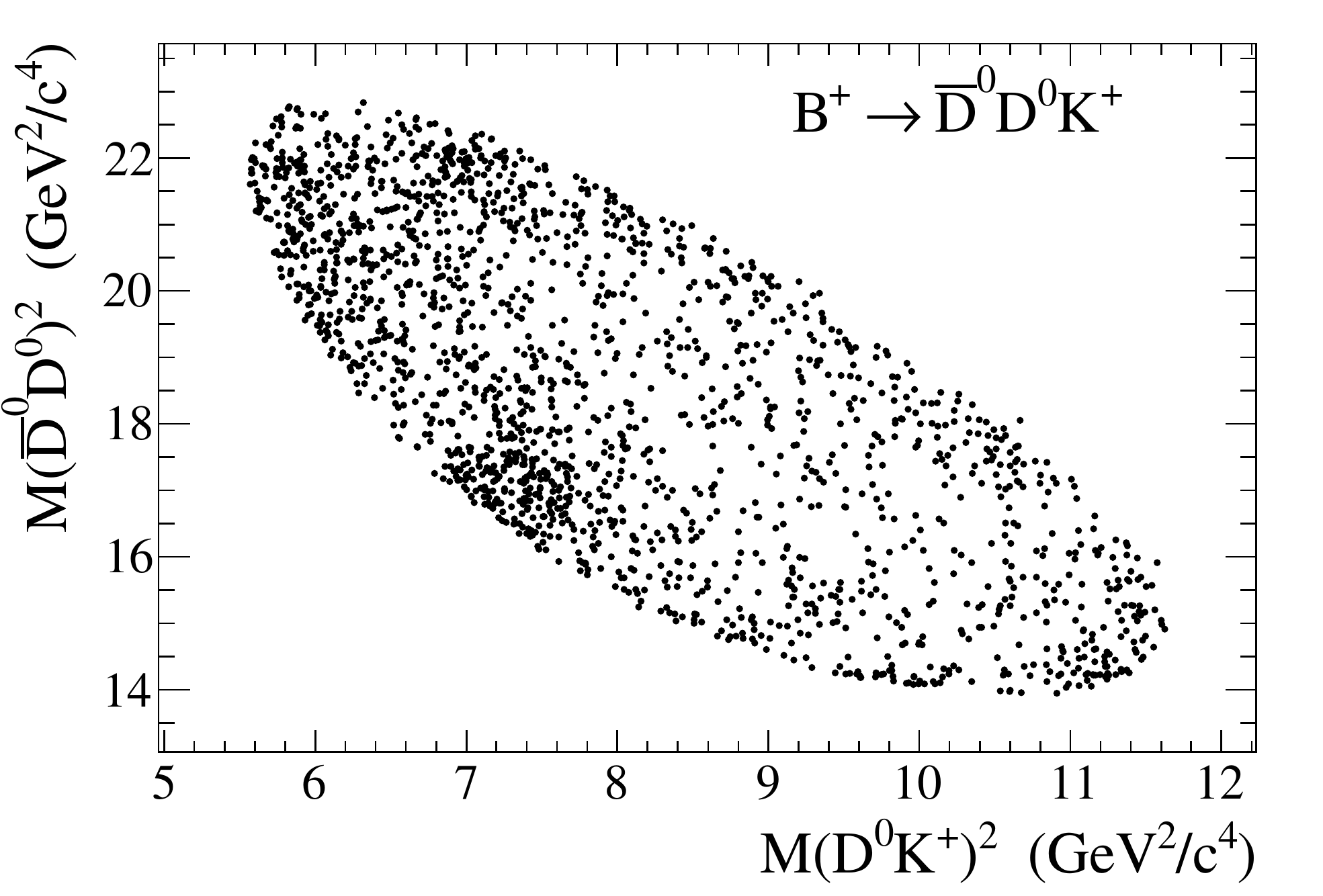,width=0.45\textwidth}
 \caption{Dalitz plots for the modes \modei (left) and \modexi (right).}
 \label{fig:dalitz}
\end{figure}

\section{Dalitz plot analyses}

\noindent
We use an isobar model formalism to perform the Dalitz plot
analysis~\cite{ref:cleoDalitz}. The formalism is well-known and we
merely note here that the dynamical amplitude, a Breit-Wigner form,
contains a multiplicative factor $F_r$ which is the Blatt-Weisskopf
damping factor for the resonance.~\cite{blatt} 

We extract the complex amplitudes present in the data (from their moduli
$\rho_i$ and phases $\phi_i$), and the mass and width of the \DsOne
resonance using an unbinned maximum likelihood fit where the minimized
negative twice-log-likelihood is called $\mathcal{F}$ and the \DsOne
resonance is the reference amplitude. We compare fits using $\Delta
\mathcal{F} \equiv \mathcal{F} - \mathcal{F}_\mathrm{nominal}$.

The initial values of the parameters are randomized and we perform 250
such fits to ensure stable convergence to a global minimum. Both the
efficiency and the background are parameterized by binned values and we
perform a 2-dimensional interpolation to obtain values at desired
locations on the plot.

The Dalitz plots for \modei and \modexi are shown in
Fig.~\ref{fig:dalitz}. The known amplitudes that could give a
contribution in the Dalitz plot for both modes are nonresonant events,
the \DsOne meson, and the \DsTwo meson, which can decay to $D^0K^+$, but
has not been observed in $B \to \Dbar^{(*)} D K$ decays.  The \DsJ{2860}
state is not included in the nominal fit. For the mode \modexi, the
additional contributions from charmonium states included in the fit are
the \PsiRes meson, and the $\psi(4160)$ meson.

In the following fits, the masses and widths of these resonances are
fixed to their world averages~\cite{ref:pdg}, except for the \DsOne
where the parameters are free to vary. The spin of this resonance is
assumed to be 1.

\noindent
Preliminary fits with the components mentioned above fail to give
satisfactory $\chi^2/\ndof$. Also, the low mass region between 2350 and
2500 $\mevcc$ is not well described, especially for \modexi.

\begin{figure*}[htb]
\centering
 \epsfig{file=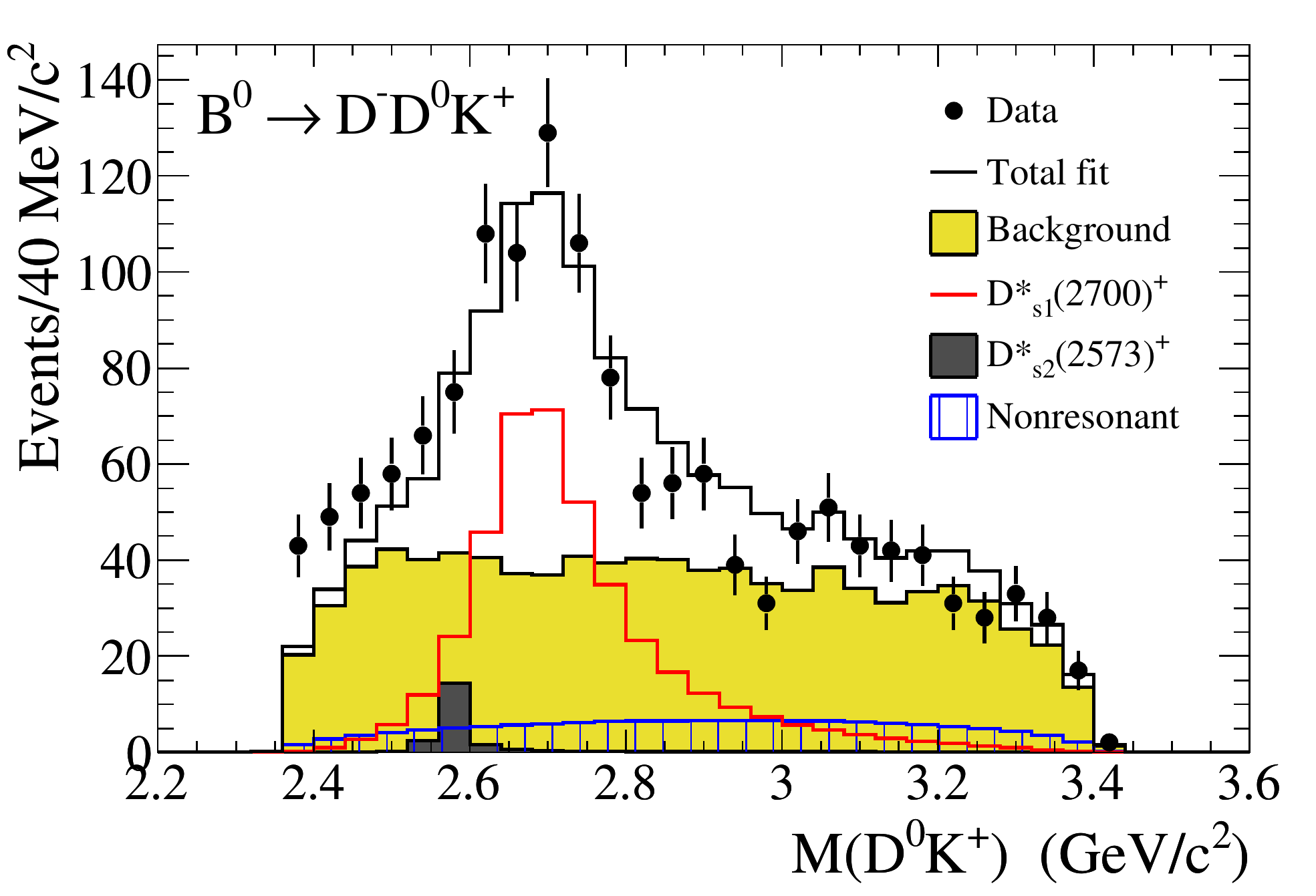,width=0.45\textwidth}
 \epsfig{file=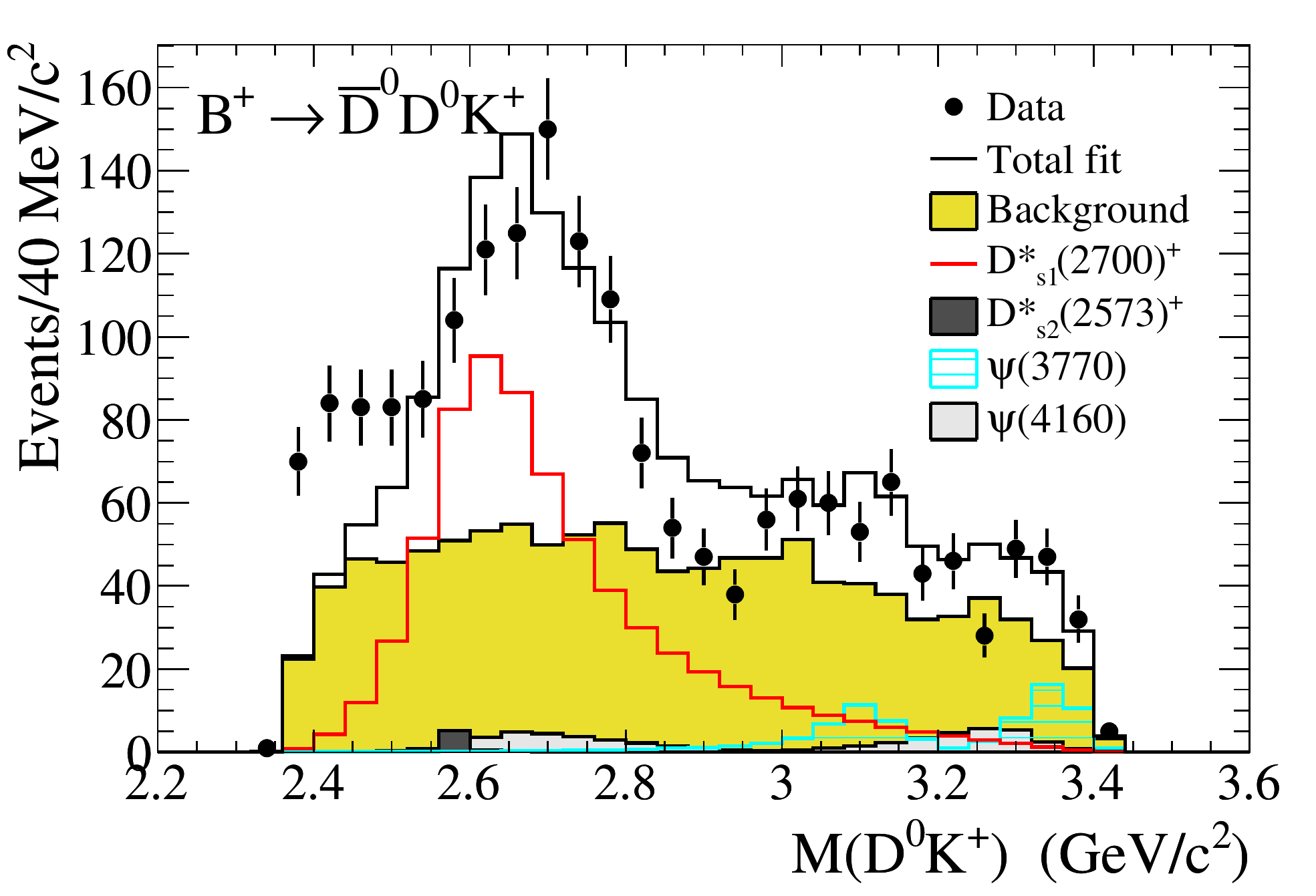,width=0.45\textwidth}
 \caption{Projections of the Dalitz plot on the $D^0K^+$ axis for the
   data (dots) and for the result of the preliminary fit (total
   histogram) for the modes \modei (left) and \modexi (right).}
 \label{fig:dal_nolow}
\end{figure*}

\begin{figure*}[htb]
\centering
 \epsfig{file=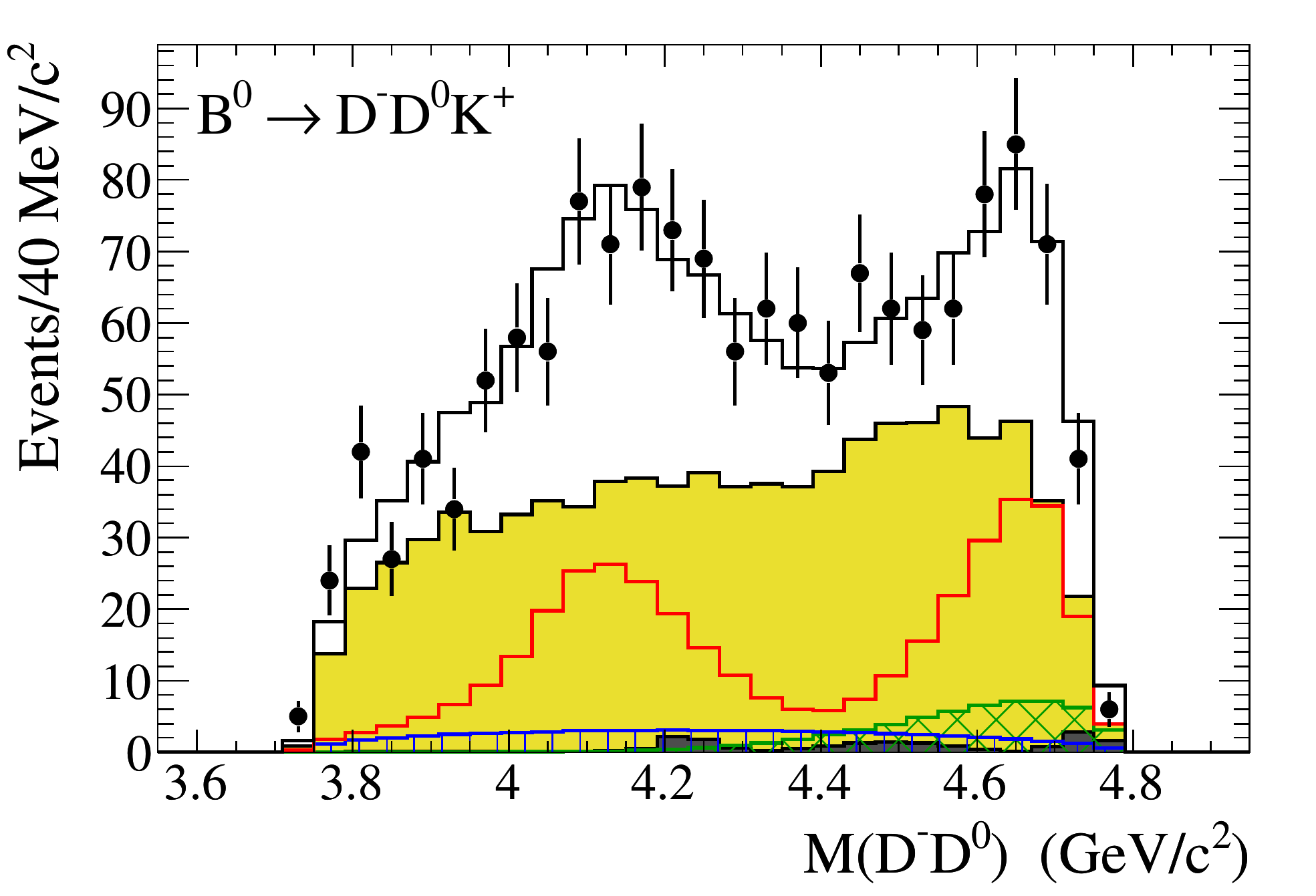,width=0.45\textwidth}
 \epsfig{file=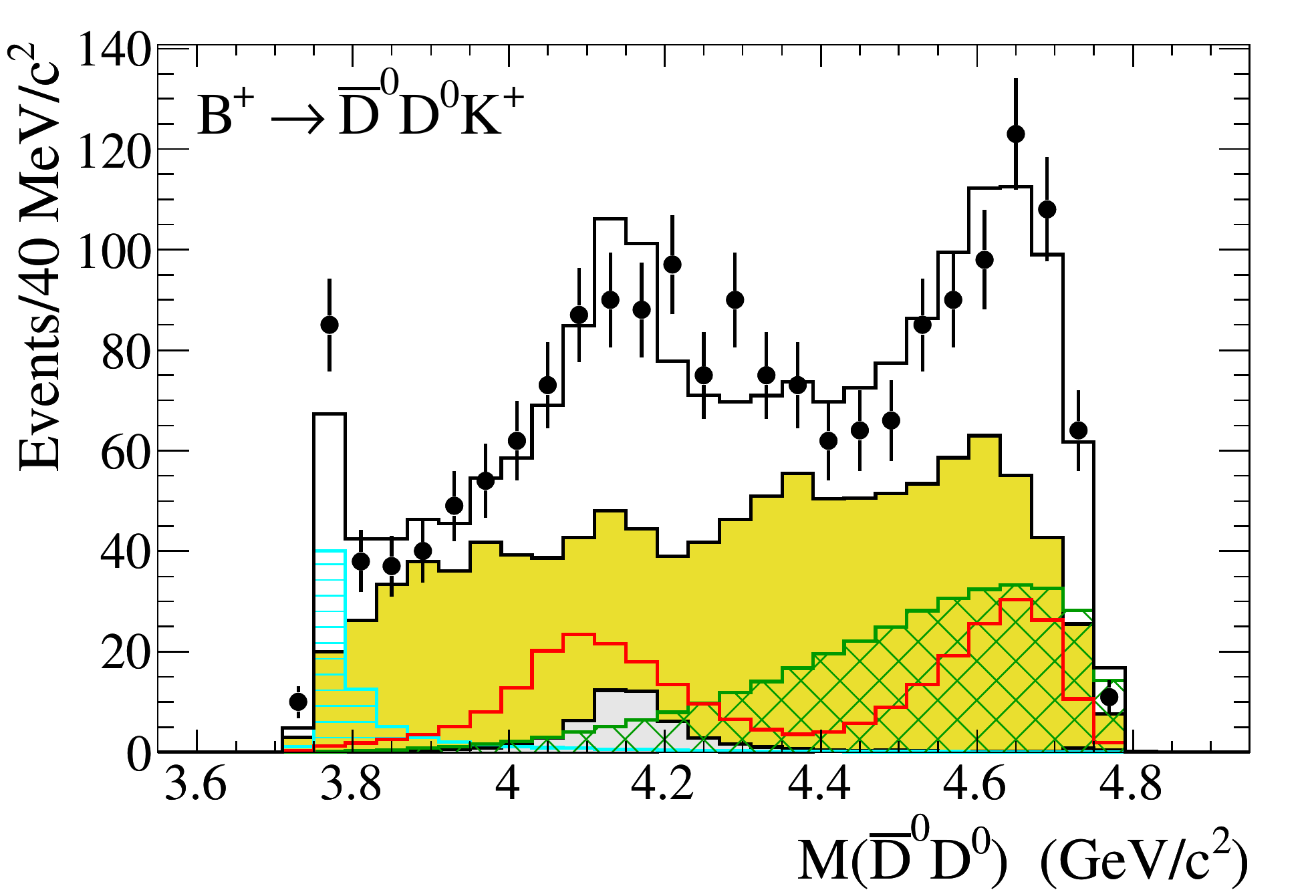,width=0.45\textwidth}
 \caption{Projections of the Dalitz plot on $M(D\overline{D})$ axis for the data
   (dots) and for the result of the nominal fit (total histogram) for
   the modes \modei (left) and \modexi (right).}
 \label{fig:dal_dbard}
\end{figure*}

\begin{figure*}[htb]
\centering
 \epsfig{file=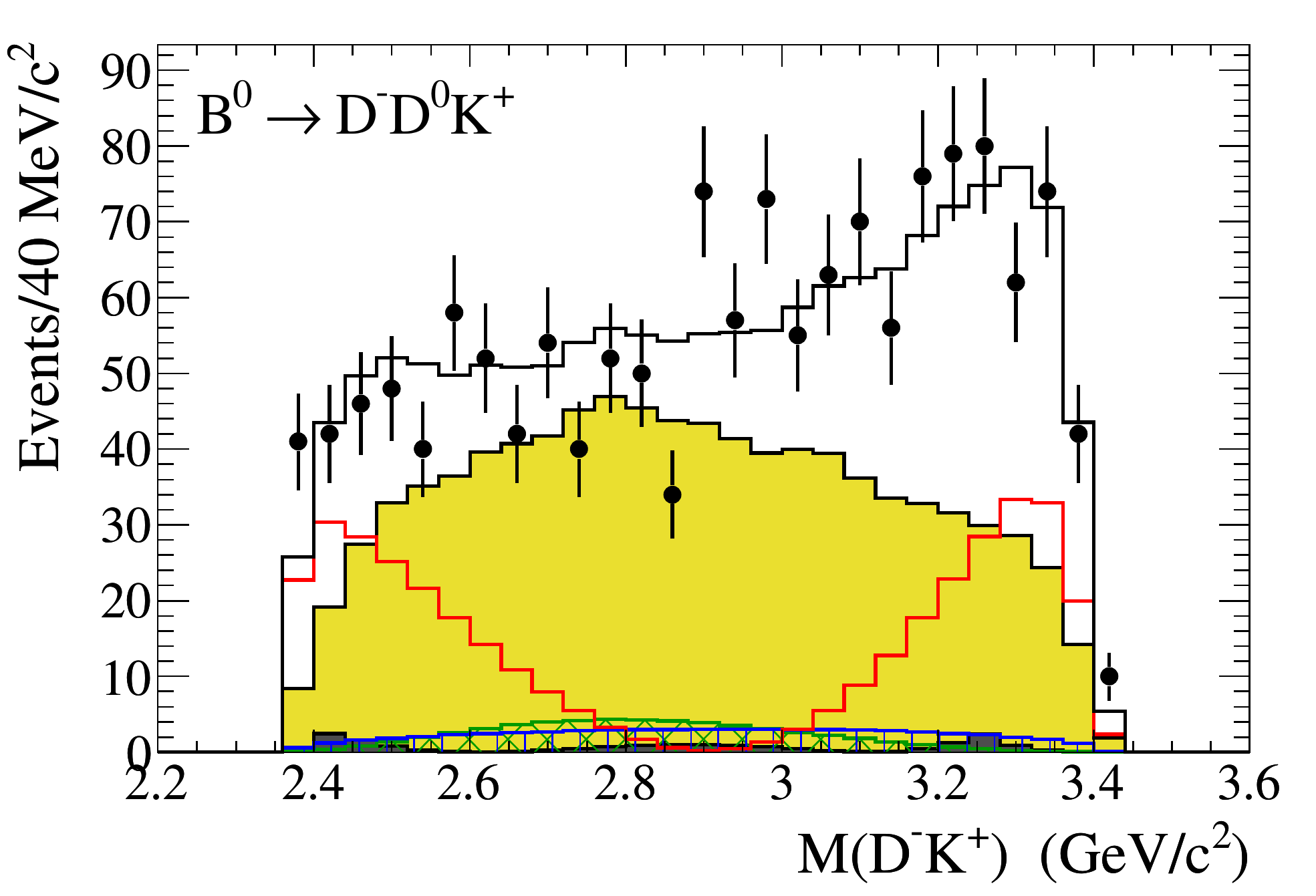,width=0.45\textwidth}
 \epsfig{file=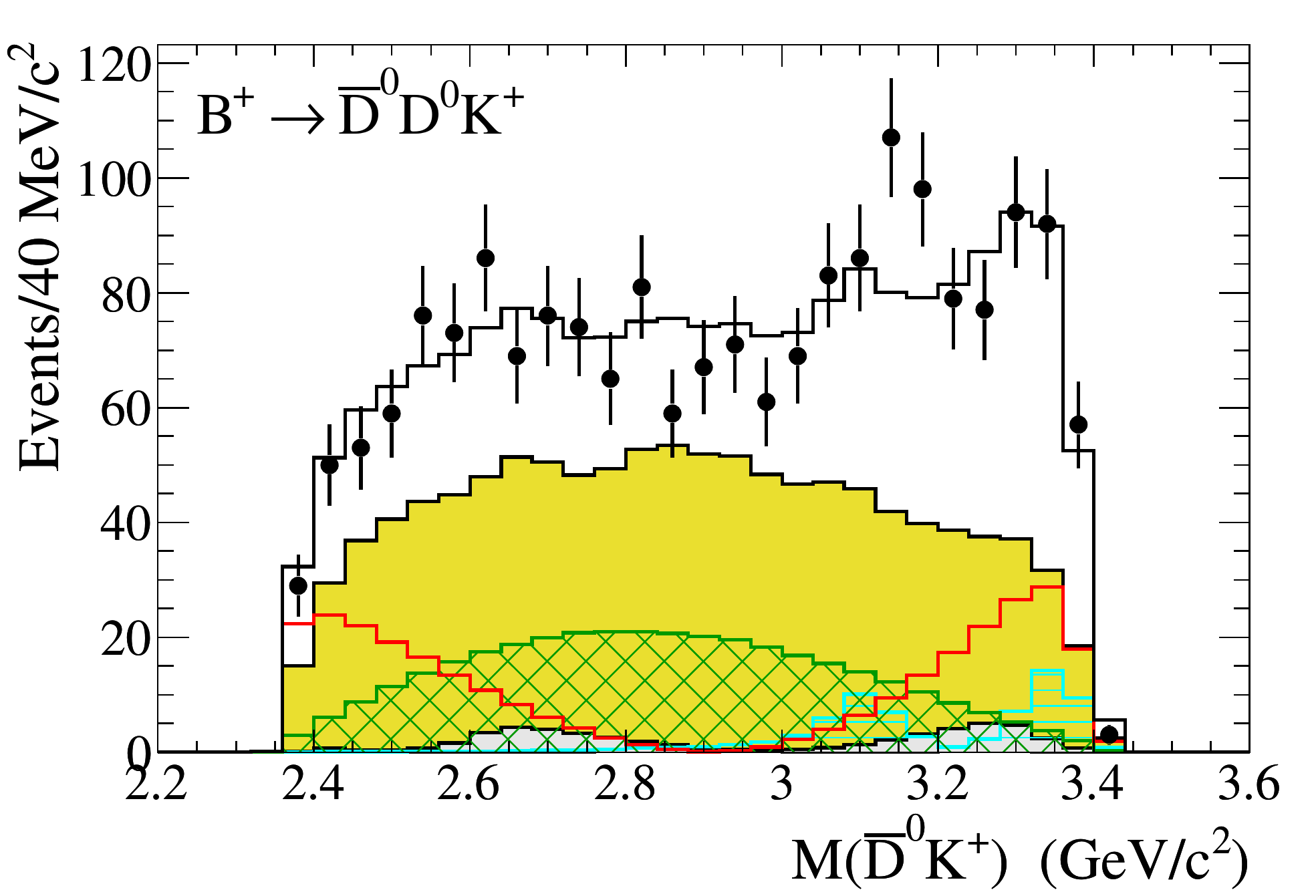,width=0.45\textwidth}
 \caption{Projections of the Dalitz plot on $M(\overline{D}K)$ axis for the data
   (dots) and for the result of the nominal fit (total histogram) for
   the modes \modei (left) and \modexi (right).}
 \label{fig:dal_dbark}
\end{figure*}

\begin{figure*}[htb]
\centering
 \epsfig{file=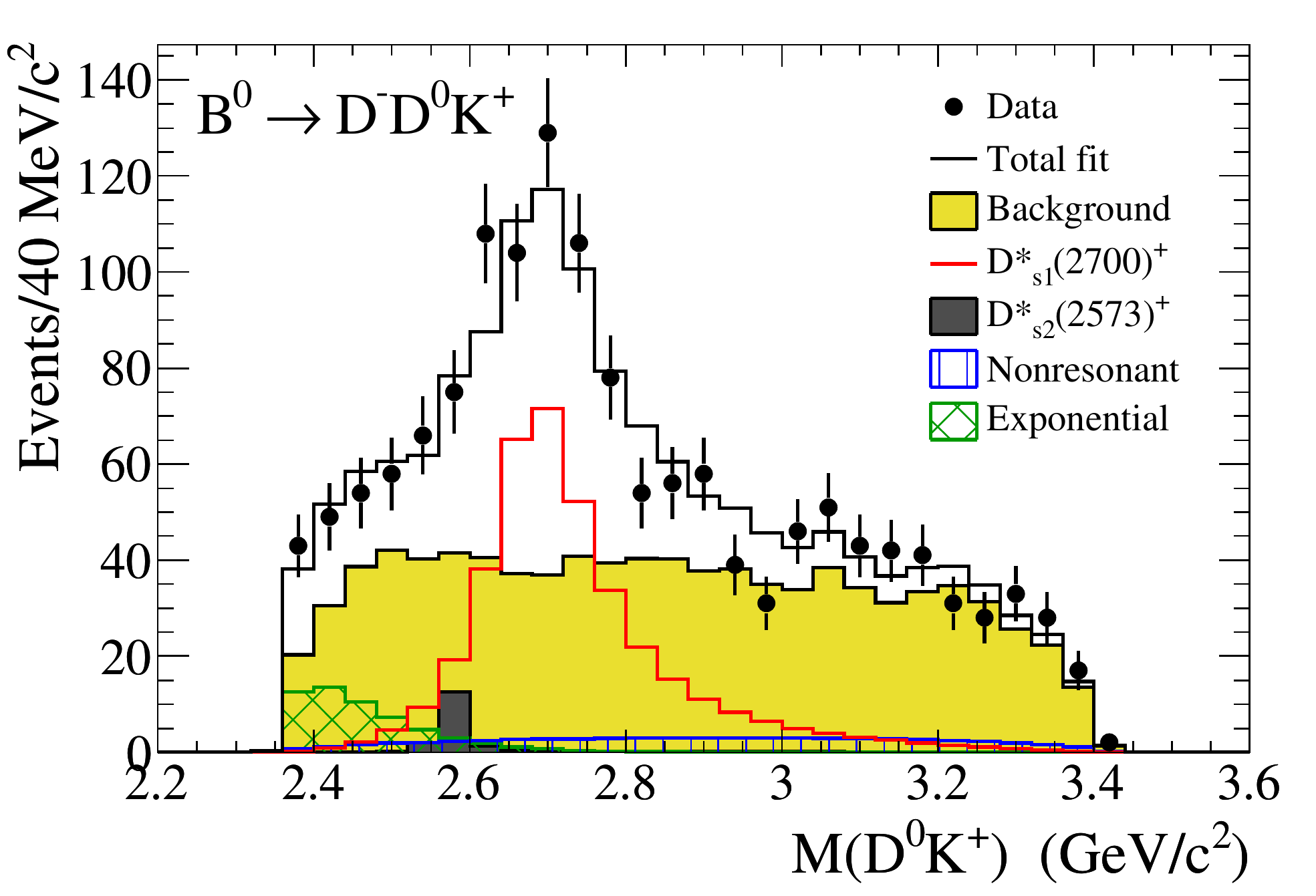,width=0.45\textwidth}
 \epsfig{file=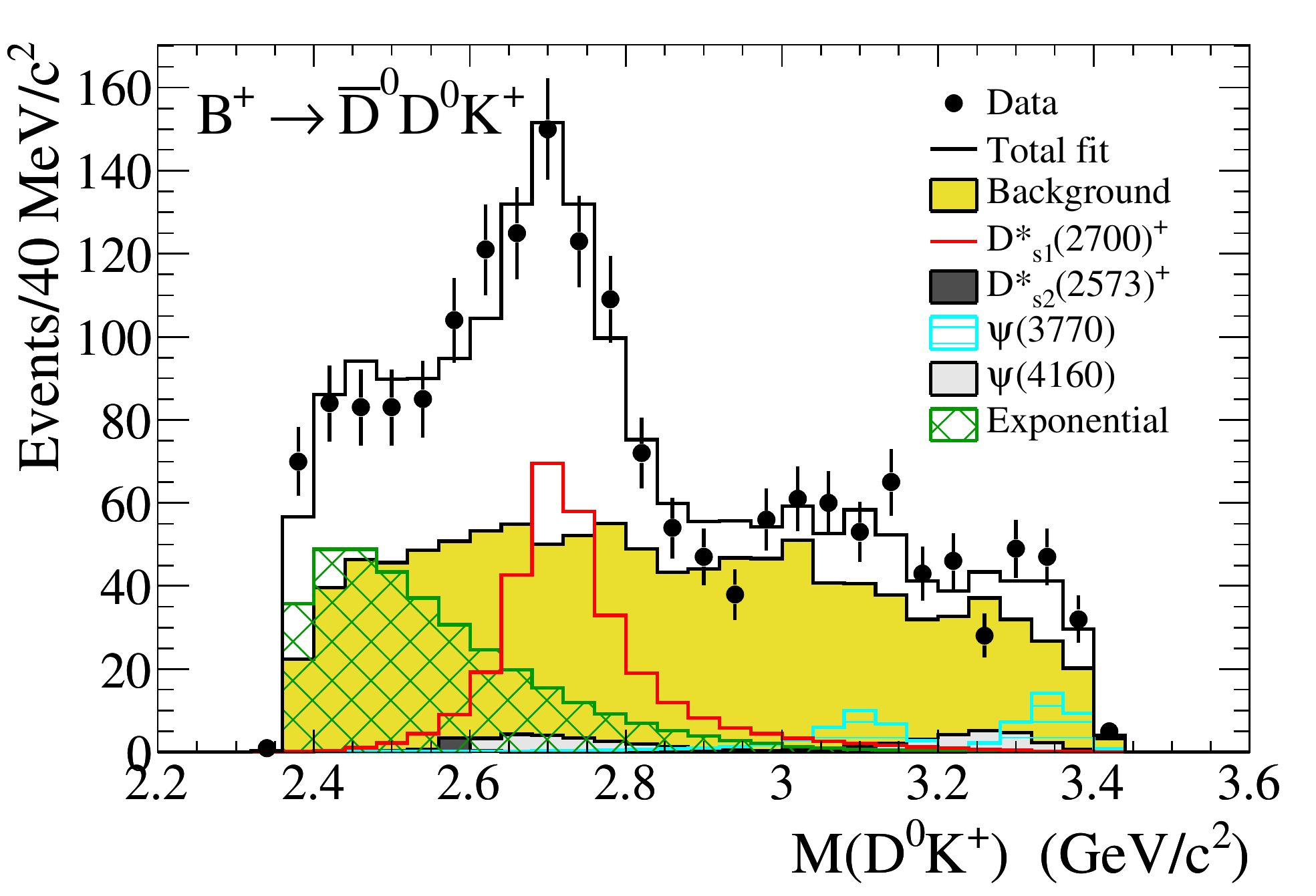,width=0.45\textwidth}
 \caption{Projections of the Dalitz plot on $M(DK)$ axis for the data
   (dots) and for the result of the nominal fit (total histogram) for
   the modes \modei (left) and \modexi (right).}
 \label{fig:dal_dk}
\end{figure*}

After verifying that this enhancement is due to signal and not
background or a cross-feed reflection, using two methods to subtract
background, we proceed to fit it using an ad-hoc function that
exponentially falls with $m^2(D^0K^+)$. We call these our nominal
fits. Since this region does not overlap significantly with the \DsOne
region of interest, we are content to ascribe a systematic error to this
effect. [We tried a low mass resonance but it only marginally improves
  the nominal fit.] The nominal fit for \modei returns
$\chi^2/\ndof=56/45$ and the nominal fit for \modexi gives
$\chi^2/\ndof=86/48$. These fits are presented in
Figures~\ref{fig:dal_dbard}, ~\ref{fig:dal_dbark}, and
~\ref{fig:dal_dk}. The high value of the $\chi^2/\ndof$ can be partly
explained by differences in the data and fit densities in the \PsiRes
region.

We consider several sources of systematic uncertainties in the fit
parameters such as the moduli, the phases, the fit fractions, and the
mass and width of the \DsOne; the combined result of this investigation
is reported as the second error below.

\section{Results}

\noindent
The results for the Dalitz plot analysis of the modes \modei and \modexi
are presented in Tables~\ref{tab:result1} and~\ref{tab:result11}. This
is the first time the \DsOne is observed in the decay \modei. An excess
at low $D^0K^+$ invariant mass is evident but we have been unable to
determine its origin.

\begin{table*}[htb]
 \begin{center}
 \caption{Results from the Dalitz plot fit (moduli, phases, and
   fractions) for \modei. The different contributions are listed: the
   \DsOne and \DsTwo resonances, the nonresonant amplitude and the
   low-mass excess described by an exponential. The first uncertainties
   are statistical and the second systematic.}
 \label{tab:result1} \vskip 0.2cm
 \begin{tabular}{lrrr}
 \hline
 \hline
 \CellTop
Contribution & Modulus & Phase ($^\circ$) & ~Fraction (\%) \\
\hline
\CellTopTwo
\DsOne & 1.00 & 0 & $66.7 \pm 7.8 {}^{+3.5}_{-3.8}$ \\
\CellTopTwo
\DsTwo & $0.031 \pm 0.008 \pm 0.002$ & $277 \pm 17 {}^{+6}_{-9}$ & $3.2 \pm 1.6 {}^{+0.3}_{-0.4}$ \\
\CellTopTwo
Nonresonant & $1.33 \pm 0.63 {}^{+0.46}_{-0.35}$ & $287 \pm 21 {}^{+10}_{-15}$ & $10.9 \pm 6.6 {}^{+7.0}_{-4.3}$ \\
\CellTopTwo
Exponential & $6.94 \pm 1.83 {}^{+0.82}_{-0.43}$ & $269 \pm 33 {}^{+17}_{-15}$ & $9.9 \pm 2.9 {}^{+3.0}_{-3.3}$ \\
\hline
\CellTopTwo
Sum & & & $90.6 \pm 10.7 {}^{+8.4}_{-6.7}$ \\
\hline
\hline
 \end{tabular}
 \end{center}
\end{table*}

\begin{table*}[htb]
 \begin{center}
 \caption{Results from the Dalitz plot fit (moduli, phases, and
   fractions) for \modexi. The different contributions are listed: the
   \DsOne, \DsTwo, \PsiRes, and $\psi(4160)$ resonances, and the
   low-mass excess described by an exponential. The first uncertainties
   are statistical and the second systematic.}
 \label{tab:result11} \vskip 0.2cm
 \begin{tabular}{lrrr}
 \hline
 \hline
 \CellTopTwo
Contribution & Modulus & Phase ($^\circ$) & ~Fraction (\%) \\
 \hline
 \CellTopTwo
\DsOne & 1.00 & 0 & $38.3 \pm 5.0 {}^{+0.8}_{-6.2}$ \\
 \CellTopTwo
\DsTwo & $0.021 \pm 0.010 {}^{+0.009}_{-0.003}$ & $267 \pm 30 {}^{+17}_{-13}$ & $0.6 \pm 1.1 {}^{+0.4}_{-0.2}$ \\
 \CellTopTwo
\PsiRes & $1.40 \pm 0.21 {}^{+0.20}_{-0.24}$ & $284 \pm 22 {}^{+26}_{-30}$ & $9.0 \pm 3.1 {}^{+0.4}_{-0.8}$ \\
 \CellTopTwo
$\psi(4160)$ & $0.78 \pm 0.20 {}^{+0.18}_{-0.14}$ & $188 \pm 13 {}^{+14}_{-17}$ & $6.4 \pm 3.1 {}^{+1.9}_{-2.4}$ \\
 \CellTopTwo
Exponential & $16.15 \pm 2.26 {}^{+1.09}_{-1.74}$ & $308 \pm 8 {}^{+6}_{-5}$ & $44.5 \pm 6.2 {}^{+1.3}_{-2.1}$ \\
\hline
 \CellTopTwo
Sum & & & $98.9 \pm 9.2 {}^{+2.5}_{-7.0}$ \\
\hline
\hline
 \end{tabular}
 \end{center}
\end{table*}

Using the Dalitz fit fractions and the total branching fractions
measured in a previous publication~\cite{ref:DDKBF} with the exact same
data sample we obtain the results presented in
Table~\ref{tab:summaryBF}.

\begin{table}[htb]
 \begin{center}
 \caption{Summary of partial branching fractions. The first
   uncertainties are statistical and the second systematic. The notation
   $B^{0} \to D^{-} \DsOne \; [D^{0}K^{+}]$ refers to $B^{0} \to D^{-}
   \DsOne$ followed by $\DsOne \to D^{0}K^{+}$.}
 \label{tab:summaryBF}
  \vskip 0.2cm
 \begin{tabular}{ll}
 \hline
 \hline
 \CellTop
 Mode & \cal{B} \, $(10^{-4})$ \\
 \hline
 \CellTop
$B^{0} \to D^{-} \DsOne \; [D^{0}K^{+}]$ ~~~~~~~& $7.14 \pm 0.96 \pm 0.69$\\
$B^{+} \to \Dbar^{0} \DsOne \; [D^{0}K^{+}]$ & $5.02 \pm 0.71 \pm 0.93$\\
$B^{0} \to D^{-} \DsTwo \; [D^{0}K^{+}]$ & $0.34 \pm 0.17 \pm 0.05$\\
$B^{+} \to \Dbar^{0} \DsTwo \; [D^{0}K^{+}]$ & $0.08 \pm 0.14 \pm 0.05$\\
$B^{+} \to \PsiRes K^{+} \; [\Dbar^{0}D^{0}]$ & $1.18 \pm 0.41 \pm 0.15$\\
$B^{+} \to \psi(4160) K^{+} \; [\Dbar^{0}D^{0}]$ & $0.84 \pm 0.41 \pm 0.33$\\
\hline
\hline
 \end{tabular}
 \end{center}
\end{table}

\begin{table}[htb]
 \begin{center}
 \caption{Mass and width of the \DsOne\ meson obtained from the Dalitz
   plot analyses of the modes \modei and \modexi. The first
   uncertainties are statistical and the second systematic.}
 \label{tab:allDsJ}
  \vskip 0.2cm
 \begin{tabular}{lll}
 \hline
 \hline
 \CellTop
Mode~~~~~~~~~~~~~~~~~ & Mass (MeV/$c^2$)~~~~ & Width (MeV) \\
 \hline
 \CellTop
 \modei & $2694 \pm 8 {}^{+13}_{-3}$ & $145 \pm 24 {}^{+22}_{-14}$ \\
\CellTopTwo
\modexi & $2707 \pm 8 \pm 8$ & $113 \pm 21 {}^{+20}_{-16}$ \\
\hline
\hline
 \end{tabular}
 \end{center}
\end{table}

We list results for the mass and width of the \DsOne\ meson in
Table~\ref{tab:allDsJ}. Combining modes we obtain 
\begin{eqnarray}
  M(\DsOne) &=& 2699 {}^{+14}_{-7} \mevcc,  \\
  \Gamma(\DsOne) &=& 127 {}^{+24}_{-19} MeV, \nonumber
\end{eqnarray}
compatible with the world averages. Repeating our fits with the $J=0$
and $J=2$ hypotheses (see results in Table~\ref{tab:spin}) we conclude
that $J=1$ is strongly favored; further assuming parity conservation in
the resonant decays we deduce that the \DsOne\ has $J^P=1^-$.

\begin{table*}[htb]
 \begin{center}
 \caption{Value of $\Delta \mathcal{F}$ and $\chi^2/\ndof$ for the
   hypotheses $J=0,1,2$ for the two modes. The nominal fit is presented
   in bold characters.}
 \label{tab:spin}
  \vskip 0.2cm
 \begin{tabular}{lllclll}
 \hline
 \hline
 \CellTop
Mode & \multicolumn{2}{c}{$J=0$} & \multicolumn{1}{c}{$J=1$} & \multicolumn{2}{c}{$J=2$} \\
& $\Delta \mathcal{F}$~~ & $\chi^2/\ndof$ & ~~~~~~$\chi^2/\ndof$~~~~~~ & $\Delta \mathcal{F}$~~ & $\chi^2/\ndof$ \\
 \hline
 \CellTop
\modei ~~~~~~~~ & $131$ & $131/45$ & \textbf{56/45} & $108$ & 125/45 \\
\CellTop
\modexi & $63$ & 137/48 & \textbf{86/48} & $99$ & 145/48 \\
\hline
\hline
 \end{tabular}
 \end{center}
\end{table*}

Finally, our fits do not favor inclusion of the \DsJ{2860} and
\DsJ{3040} resonances in the final states \modei and \modexi. 

\Acknowledgements
We would like to acknowledge the efforts of the \babar collaboration.

\end{document}

%% file: packages.tex
\usepackage{dcolumn}
\usepackage{graphicx}
\usepackage{lineno}
\usepackage[allowmove]{url}
\usepackage{verbdef}
\usepackage{xstring}
\usepackage{xspace}
\usepackage{ifthen}
\usepackage{xifthen}
\usepackage{etoolbox}
\usepackage{hyperref}
\usepackage{enumitem}
\usepackage{makecmds}
\usepackage{amsmath}
\usepackage{caption}
\usepackage{subcaption}

\usepackage{epsfig}
\usepackage{verbatim}
\usepackage{mathrsfs}

\usepackage{color}

%% file: econfmacros.tex
\def\beq{\begin{equation}}
\def\eeq#1{\label{#1}\end{equation}}
\def\eeqn{\end{equation}}

\def\beqa{\begin{eqnarray}}
\def\eeqa#1{\label{#1}\end{eqnarray}}
\def\eeqan{\end{eqnarray}}

\let\bar=\overbar

\def\Dslash{\not{\hbox{\kern-4pt $D$}}}
\def\dslash{\not{\hbox{\kern-2pt $\del$}}}

\def\msb{{\bar{\ssstyle M \kern -1pt S}}}

